\documentclass[prx,aps,showpacs,twocolumn,preprintnumbers,amsmath,amssymb,longbibliography]{revtex4-1}
\usepackage[english]{babel}
\usepackage{amsmath,amssymb,bbm,mathrsfs,mathtools,multirow,diagbox,xcolor,%
  graphicx,tabularx,comment,amsfonts,dsfont,lettrine,units,comment}
\usepackage{graphicx}
\usepackage[colorlinks=True,linkcolor=red,citecolor=blue,urlcolor=blue]{hyperref}
\usepackage{bookmark}
\usepackage{xcolor}
\usepackage{chngcntr}
\usepackage{braket}
\usepackage{nicefrac}
\usepackage{bm}
\usepackage{bbm}
\usepackage{braket}
\usepackage{lineno}
\usepackage{pdfpages}
\usepackage{array}
\newcolumntype{C}{>{$}c<{$}}
\usepackage{newtxtext} 
\usepackage{ulem}   

\newcommand{\diff}{\mathrm{d}}
\def\br{\mathbf{r}}
\def\bk{\mathbf{k}}

\def\ket#1{|#1\rangle }
\def\bra#1{\langle #1 |}

\makeatletter
\AtBeginDocument{\let\LS@rot\@undefined}
\makeatother

\usepackage{xstring} 

\newcommand{\customSection}[1]{{{\it{#1.}}---}}

\begin{document}

\title{Enhanced quantum metric due to vacancies in graphene}

\author{Quentin Marsal}
 \affiliation{Department of Physics and Astronomy, Uppsala University, Box 516, 
751 20 Uppsala, Sweden}

\author{Annica M. Black-Schaffer}
 \affiliation{Department of Physics and Astronomy, Uppsala University, Box 516, 
751 20 Uppsala, Sweden}

\begin{abstract}
    Random vacancies in a graphene monolayer induce defect states are known to from a narrow impurity band centered around zero energy at half-filling.  
    We use a space-resolved formulation of the quantum metric and establish a strong enhancement of the electronic correlations in this impurity band. 
    The enhancement is primarily due to strong correlations between pairs of vacancies situated on different sublattices at anomalously large spatial distances.     
     We trace the strong enhancement to both the multifractal vacancy wave functions, which ties the system exactly at the Anderson insulator transition for all defect concentrations, and preserving the chiral symmetry.
\end{abstract}

\date{\today}

\maketitle
The two-dimensionality and linear Dirac dispersion of graphene generate remarkable physical properties \cite{novoselov2004, novoselov2005, castro2009, wehling2014}. However, the vanishing density of states at half-filling, at the Dirac point, causes a natural insensitivity to electron interactions and its generated many-body instabilities, otherwise key concepts in condensed matter. But, it makes graphene sensitive to the presence of structural defects~\cite{castro2009, banhart2011, yang2018}. Defects often create localized states that affects the conductivity, including in graphene~\cite{stauber2007, wehling2010}, and can even turn a metal into an insulator~\cite{evers2008}, while defect states that are not fully localized have been shown to favor some long-range correlations~\cite{feigelman2007, burmistrov2012, burmistrov2021}.

In this work, we consider the archetype defect in graphene: vacancies~\cite{ugeda2010, ugeda2011}. 
Vacancies, as well as hydrogenation~\cite{banhart2011, yang2018} which induce the same type of lattice defects, are both common and controllable.
Graphene vacancies create critical defect states pinned at the zero-energy Dirac point by the chiral symmetry~\cite{evers2008, gade1993, altland1999, fukui2000, ostrovsky2006, ostrovsky2007,pereira2006,ramires1019}.
However, the disorder-origin of the resulting zero-energy flat band raises a strong initial hesitation about its ability to generate physics beyond the localized disorder regime, including many-body instabilities or transport properties. 

We use a space-resolved version of the Fubini-Study quantum metric~\cite{marzari1997} and show that vacancies in graphene introduce strong long-range correlations.
In flat bands in crystalline systems, such a finite quantum metric has recently been shown to generate superconducting phase coherence~\cite{peotta2015, peotta2023, julku2016, Kopnin2011}, as well as being needed to assess the existence of fractional excitations~\cite{varjas2022}. 
This includes the plethora of phenomena recently discovered in magic-angle twisted or strongly biased bi- and trilayer graphene due to its narrow crystalline flat bands \cite{Cao2018SC,Hao2021, Park2021, Zhou2021SC, zhou2022, kim2022, andrei2020}. 
A finite quantum metric has also recently been shown to give a nonlinear anomalous Hall effect \cite{Gao2023, Wang2023}.
Our results establish that these discovered phenomena in crystalline systems can also be accessible by simply using localized disorder.

\customSection{Space-resolved quantum metric} 
\label{sec:Qmetric}
In crystalline systems, due to translational invariance, Bloch's states $u_k^\alpha$ form an eigenbasis for the Hamiltonian, with $\alpha$ labeling bands and $k$ the momentum of the quantum state. 
For a given filling fraction, the quantum metric then reads~\cite{peotta2015, marzari1997}:
\begin{equation}
    \Omega_{\mu\nu} = \int_{BZ} \diff\bk \sum_{\alpha\ \mathrm{filled}} \left(\partial_{k_\mu}u_k^{\alpha}\right)^\dagger (\mathbb{I}-u_k^\alpha u_k^{\alpha\dagger})\left(\partial_{k_\nu}u_k^\alpha\right) \label{eq:QM_k},
\end{equation}
with the indices $\mu,\nu$ enumerating the spatial dimensions.
The quantum metric measures the distance in Hilbert space between states corresponding to neighboring momenta $k$ and $k + \diff k$. With the integration over the Brillouin zone, the quantum metric is the total volume spanned by the filled band states in the Hilbert space.

With disorder present, a real space formulation of the quantum metric is instead required. The origin of such can be traced back to work defining maximally localized Wannier functions \cite{marzari1997, resta1999,peotta2023,tam2023}. From there the quantum metric can be identified as the reverse Fourier transform of Eq.~\eqref{eq:QM_k} \cite{marzari1997}:
\begin{equation}
    \Omega_{\mu\nu} = \mathrm{Tr}\left[P\br_\mu (1-P) \br_\nu P\right]\label{eq:QM_r},
\end{equation}
with multiplications by the position operator $\br$ replacing momenta derivatives and $P$ being the projector on the occupied states. 
Due to the trace, $\Omega_{\mu\nu}$ can be computed over any basis of the Hilbert space. 
In crystalline systems, the usual basis is the Bloch's states, as in Eq.~\eqref{eq:QM_k}. 
In non-crystalline systems we find it useful to compute the trace in the site (atomic) basis, $\ket{i}$, to be able to define a space-resolved quantum metric marker: $\bra{i}P\br_\mu (1-P) \br_\nu P\ket{i}$. 
Together with the recently developed local Chern marker~\cite{bianco2011, marsal2020}, this provide a fully space-resolved notion of the quantum geometric tensor~\cite{resta2002}.

A physical interpretation of Eq.~\eqref{eq:QM_r} is that it characterizes the spreading of the occupied states, as $\Omega = \Omega_{xx} + \Omega_{yy}$ can be rewritten as the expectation value $\left<\br^2\right>-\left<\br\right>^2$~\cite{marrazzo2019, marzari1997, resta1999}. 
In particular, focusing on a single band, the functional $\Omega$ is measuring the quadratic spreading of the maximally localized Wannier orbitals that can be built for that band. 
As a consequence, if the Wannier functions can be fully localized around single atomic sites, their spreading is practically zero and the quantum metric vanishes, defining a trivial atomic insulator. 
In contrast, in topologically non-trivial bands, the Wannier basis cannot be fully localized~\cite{bradlyn2017}, and the quantum metric is bounded by below by the topological invariant,  e.g.~in two-dimensional Chern insulators $\Omega \geq \frac{A}{2\pi}|C|$, with $C$ the Chern number and A the sample area. \cite{peotta2015}.

\begin{figure}
    \centering
    \includegraphics[width = \columnwidth]{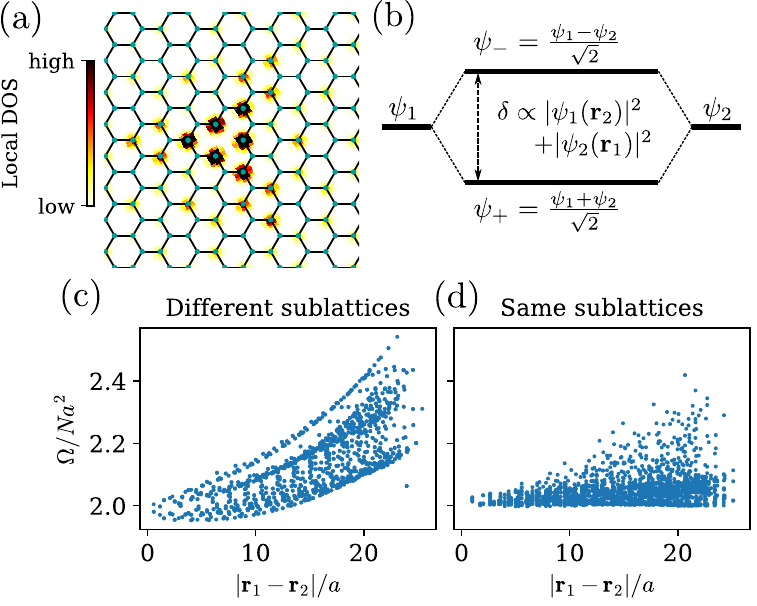}
\caption{(a) Local density of states at zero energy for a single vacancy. The defect wave function is localized on the opposite sublattice to that of the vacancy. (b) Two vacancies $\ket{\psi_{1,2}}$ on different sublattices hybridize, forming bonding $\ket{\psi_{+}}$ and anti-bonding $\ket{\psi_{-}}$ states delocalized over the two vacancies. Quantum metric (per unit cell) as a function of vacancy separation, for random vacancies pairs on different (c) and same (d) sublattice. Here $N = 1174$ for a hexagonal-shaped lattice with armchair edges, with one vacancy at the center and the other successively occupying all the other sites. }
    \label{fig:dim}
\end{figure}

\customSection{Graphene vacancies}
With a space-resolved quantum metric now at hand we apply it to show how systems with localized disorder can host a non-trivial quantum metric.
In particular, we focus on a random distribution of vacancies in a graphene monolayer. 
We model graphene using the usual tight-binding Hamiltonian on honeycomb lattice, encoding the conducting $p_z$ electron on each site interacting through nearest-neighbor hopping and with unit cell length $a$ \cite{castro2009}.
We then introduce vacancies by removing a number $N_{vac}$ of sites $(v_i)$ from the lattice, setting all hopping terms to the vacancy site to zero, resulting in the Hamiltonian
\begin{equation}
    \mathcal{H} = \sum_{\left<i,j\right>}t c_i^\dagger c_j - \sum_{\left<v_i,j\right>}t (c_{v_i}^\dagger c_j + H.c.).
    \label{eq:H}
\end{equation}
Finally, we remove the trivially isolated orbitals $c_{v_i}$ from the Hilbert space. We typically consider large systems with open boundary conditions to avoid periodic images, using up to $N=5400$ sites and check our key results for spurious finite size effects.
As a vacancy defect only acts on bonds (by removing them), it preserves the chiral symmetry of graphene and is thus a type of chiral disorder, which has been studied both generically and in the honeycomb lattice \cite{evers2008, gade1993, altland1999, fukui2000, ostrovsky2006, ostrovsky2007}.
Each vacancy removes one state from the spectrum, thereby leaving one state that has to be its own chiral partner, thus pinned to zero energy.
This defect state is localized on the opposite sublattice to that of the vacancy (see Fig.~\ref{fig:dim}a) and, often missed, it is multifractal, meaning its wave function amplitude decreases as a power-law with distance from the vacancy, in contrast to both fully localized defect states and extended Bloch states.
While carbon vacancies may also induce lattice reconstruction~\cite{meyer2008}, hydrogenation also effectively remove one $p_z$ electron and occur primarily without lattice reconstruction and is experimentally controllable ~\cite{banhart2011, yang2018}.
Technically, adatoms create an onsite perturbation, breaking chiral symmetry, but hydrogen absorbs strongly and still closely resembles a vacancy~\cite{casolo2009, gonzales2016} and can thus be modeled by Eq.~\eqref{eq:H}.

\customSection{Vacancy pair}\label{sec:twovacs}
We first illustrate the behavior of a single pair of vacancies.
The situation is particularly intriguing when the two vacancies belong to {\it different} sublattices. 
The two defect states then hybridize, forming bonding and anti-bonding states that are chiral symmetry partners, see also Supplementary Material (SM) \cite{SM}. We illustrate this in Fig.~\ref{fig:dim}(b), using $\psi_i(\br) = \psi(\br-\br_i)$  for the wave function of a vacancy at position $\br_i$ and $\ket{\psi_\pm}$ for the bonding/anti-bonding states. The hybridized vacancy states $\ket{\psi_\pm}$ are delocalized over the two vacancies and we numerically extract their energy splitting $\delta \propto |\psi_1(\br_2)|^2+|\psi_2(\br_1)|^2$.
Being delocalized, $\ket{\psi_\pm}$ must contribute to the quantum metric at half filling. Indeed, with the bonding state $\ket{\psi_+}$ occupied and the anti-bonding $\ket{\psi_-}$ not, 
their contribution to quantum metric is (for derivation see SM~\cite{SM})
\begin{equation}
\label{eq:Opair}
    \Omega_{pair} = \sum_\mu \left|\bra{\psi_+}\br_\mu\ket{\psi_-}\right|^2 = \frac{1}{4}\sum_\mu (\br_{1\mu}-\br_{2\mu})^2,
\end{equation}
as long as the vacancy states are fully rotationally symmetric, i.e.~ignoring lattice effects.
This quadratic defect contribution adds to the intrinsic quantum metric from the bulk states at half-filling, which is independent the vacancy positions. 
We illustrate this in Fig.~\ref{fig:dim}(c) showing the total quantum metric density as a function of distance separating a pair of vacancies.
The spread in $\Omega$ is due to the lattice inducing anisotropy into Eq.~\eqref{eq:Opair}  and not due to finite size effects (see SM \cite{SM}).

By contrast, no hybridization occurs when the two vacancies are on the same sublattice, since the chiral symmetry forces them to remain at zero energy.
Any superposition of these two original vacancy states thus remains a pair of eigenstates, each localized around its own vacancy, and we expect no significant contribution to the quantum metric. We verify this in Fig.~\ref{fig:dim}(d), which displays only a weak enhancement of the quantum metric and, importantly, no distance dependence. 
In fact, the quantum metric technically becomes ill-defined for same sublattice vacancies, as it is impossible to separate between occupied and unoccupied zero-energy states. Due to finite size effects we can however still often find a numerical energy separation and are thus able to produce Fig.~\ref{fig:dim}(d). We attribute the occasional larger quantum metric values to artifacts related to this numerical issue, which occasionally generates eigenstates that are superpositions of both vacancy states, giving a spurious contribution to the quantum metric.

\customSection{Vacancy band}\label{sec:vacband}
Having understood vacancy pairs, we tackle a finite distribution of vacancies. 
The spectrum of this disordered system contains a vacancy band, centered at energy $E = 0$, due to the retained chiral symmetry. If the vacancies were infinitely far from one another, chiral symmetry would pin the vacancy band exactly at zero energy, but at any finite concentration, vacancies in different sublattices hybridize in (at least) a pair-wise fashion according to the mechanism above, broadening the impurity band. 

\begin{figure}
    \centering
    \includegraphics[width = \columnwidth]{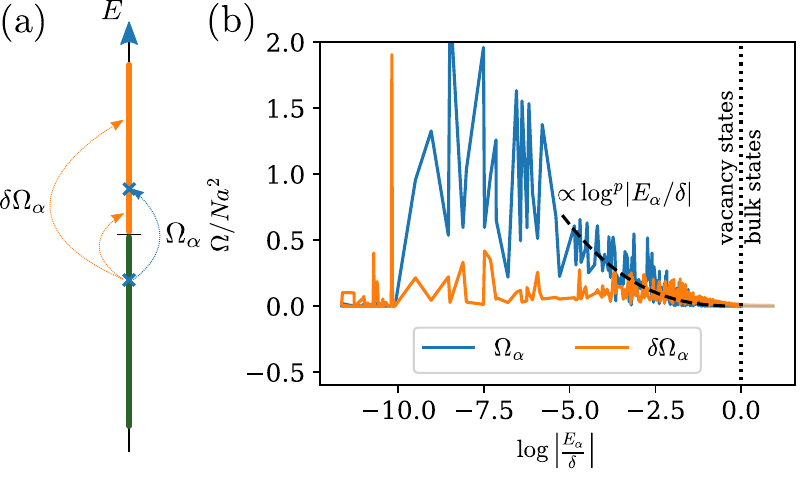}
    \caption{(a) Decomposition of the contribution to the quantum metric at half-filling from each occupied state into two parts: $\Omega_\alpha$ accounting for the correlation between the chiral symmetric states $\alpha$ and $\overline{\alpha}$ and $\delta\Omega_\alpha$ accounting for correlations with all remaining unoccupied states. (b) Contributions $\Omega_\alpha$ and $\delta\Omega_\alpha$ as a function of eigenstate energy $E_\alpha$, obtained from $20$ disorder realizations, with each configuration  $N_{vac} = 120$ in a $N = 2400$ lattice. Black dotted line separates the $N_{vac}$ closest states to zero energy, considered as vacancy states, from bulk states and defines $\delta \approx 0.34t$. Black dashed line indicates the fit described in text and the fitting range.}
    \label{fig:pair_contribution}
\end{figure}

Since eigenstates may now involve more than two vacancies, it is a priori impossible to isolate the contribution of single vacancy pairs to the quantum metric. 
Still, we can proceed because the quantum metric involves paired crossed correlations between all occupied states and all unoccupied states, see SM~\cite{SM}
\begin{equation}
    \Omega = \sum_{\mu}\sum_{\alpha\ \mathrm{filled}}\sum_{\beta\ \mathrm{empty}} \mathrm{Tr}\left[P_\alpha\br_\mu P_\beta \br_\mu P_\alpha\right],
\end{equation}
where $P_{\alpha, \beta} = \ket{\psi_{\alpha, \beta}}\bra{\psi_{\alpha, \beta}}$.
Every term of this sum is positive, no matter if the states $\alpha$ and $\beta$ are chiral symmetry partners. 
To isolate the contribution of chiral symmetric states, we denote for any occupied state $\alpha$, $\overline{\alpha}$ the (unoccupied) chiral symmetric state, i.e.~the equivalent of a vacancy pair, and use 
the decomposition
\begin{equation}
    \Omega = \sum_{\alpha\ \mathrm{filled}} \Omega_\alpha + \delta\Omega_\alpha,
\end{equation}
with $\Omega_\alpha = \mathrm{Tr}[P_\alpha\br_\mu P_{\overline{\alpha}}\br_\mu]$, and $\delta\Omega_\alpha = \mathrm{Tr}[P_\alpha\br_\mu (Q-P_{\overline{\alpha}})\br_\mu]$, with $Q = 1-P$. 
We illustrate this decomposition in Fig.~\ref{fig:pair_contribution}(a), where blue crosses illustrate a chiral symmetry pair of states.
Fig.~\ref{fig:pair_contribution}(b) shows $\Omega_\alpha$ and $\delta\Omega_\alpha$ as a function of the energy $E_\alpha$ of the eigenstate for a random distribution of vacancies.
Clearly, the strongest contribution to the quantum metric comes from the chiral symmetry partners.
In addition, we find that such partners contribute most if their energy splitting is small, corresponding to large spatial separations, in agreement with vacancy pairs in Fig.~\ref{fig:dim}(c).
This demonstrates that for a finite distribution of vacancies, there is still strong pairing between vacancies on different sublattices, with an associated large contribution to the quantum metric, especially for long-distance pairs.
We also note that bulk contributions, characterized by $|E_\alpha|$ larger than the typical impurity bandwidth $\delta$ (black dotted line in Fig.~\ref{fig:pair_contribution}(b)), are small compared to the vacancy contributions.
Moreover, if one sublattice host more vacancies than the other, they result in unpaired vacancy states at $E=0$, which, due to finite size effects, are at energies below $10^{-10}t$ and do not contribute.
If instead computing the quantum metric for finite Fermi energy $E_F$, we sum contributions from right to left, until $|E_\alpha| = |E_F|$. Then, pairs of chiral symmetric states can be both either filled or empty, which always decreases the quantum metric.

The results in Fig.~\ref{fig:pair_contribution} could in hindsight be anticipated. First, for any states, both $\Omega_\alpha$ and $\delta\Omega_\alpha$ bear a positive contribution to the quantum metric, such that there is never internal cancellations. Second, states that are not related by chiral symmetry are generally expected to be localized around different sets of vacancies, while we know that different subblatice vacancy pairs instead delocalize and generate large quantum metric contributions. 
Thus, quantum metric contributions of the form $P_\alpha\br_\mu P_\beta\br_\mu$ where $\beta\neq\overline{\alpha}$ are expected to remain smaller than contributions from chiral partners, $P_\alpha\br_\mu P_{\overline{\alpha}}\br_\mu$.
As a consequence, we find $\Omega \approx \sum_{\alpha\ \mathrm{filled}}\Omega_\alpha$.

We can further understand our results by noting that chiral disorder (disorder preserving chiral symmetry) is known to create multifractal states, meaning that they are neither exponentially localized nor fully delocalized \cite{gade1993, altland1999, fukui2000, ostrovsky2006, ostrovsky2007,hafner2014}. 
This implies a close relationship with the critical regime of the Anderson metal-insulator localization transition~\cite{evers2008}.
Indeed, vacancies in graphene, irrespective of the disorder concentration, can be expected to behave as a multifractal insulator~\cite{cuevas2007}, with enhanced correlations between states with small energy separation.
This is consistent with the enhancement of $\Omega_\alpha$ at small energies in Fig.~\ref{fig:pair_contribution}(b).
To deepen the analogy with the Anderson localization transition we write, for a pair of chiral symmetric states $\alpha$ and $\overline{\alpha}$ with energy splitting $\omega$: 
\begin{align}
    \Omega_\alpha &=\mathrm{Tr}[P_\alpha\br P_{\overline{\alpha}}\br] = \left|\bra{\psi_\alpha}\br\ket{\psi_{\overline{\alpha}}}\right|^2\\
    &= \iint \diff\br\diff\br' (\br\cdot\br')\psi_\alpha(\br)^*\psi_{\overline{\alpha}}(\br)\psi_{\overline{\alpha}}(\br')^*\psi_\alpha(\br').\label{eq:QMint}
\end{align}
$\Omega_\alpha$ is now a two-point and two-state correlator whose exact value depends on the exact position of the vacancies.
By utilizing a variational approach developed for Anderson localization \cite{houghton1980}, we find that, on average, the strongest contribution to the integral in Eq.~\eqref{eq:QMint} comes from pairs of vacancies and only depends on the energy splitting $\omega$ such that
$\Omega_\alpha\propto\log^{d+1}\left(\frac{\delta}{\omega}\right)$,
where $d = 2$ is the system dimension.
This variational approach is valid for small energy separation $\omega\ll\delta$ \cite{houghton1980, feigelman2010}, which is the case for a vacancy band.
In Fig.~\ref{fig:pair_contribution}(b) we also include a fit to the $\log^p$ functional form, finding $p = d+1 \approx 2.6$, which gives good agreement.
For this fit we exclude vacancies with $\Omega_\alpha>\frac{\sqrt{3}}{4}Na^2$, the area of the sample, which sit close to the sample edge and thus result in strong fluctuations in $\Omega_\alpha$ due to finite size effects.
These arguments establish that the strong quantum metric enhancement we observe is a consequence of the multifractality of hybridized vacancy states. 
Notably, the quantum metric enhancement is as strongest for vacancies far apart, a seemingly non-intuitive consequence of the multifractalness.

\customSection{Concentration dependence}\label{sec:density}
Above we established that vacancy pairs strongly enhance the quantum metric. A vacancy preferentially hybridize with its closest different sublattice vacancy neighbor. In a disordered system, more complex hybridization schemes involving more than two vacancies are in principle possible but we find only limited contributions from such to the quantum metric, as $\delta \Omega_\alpha$ is small. On the other hand, we find larger quantum metric contributions from vacancies with larger spatial separations. This dichotomy in the spatial dependence may result in an interesting concentration dependence.

\begin{figure}
    \centering
    \includegraphics[width = \columnwidth]{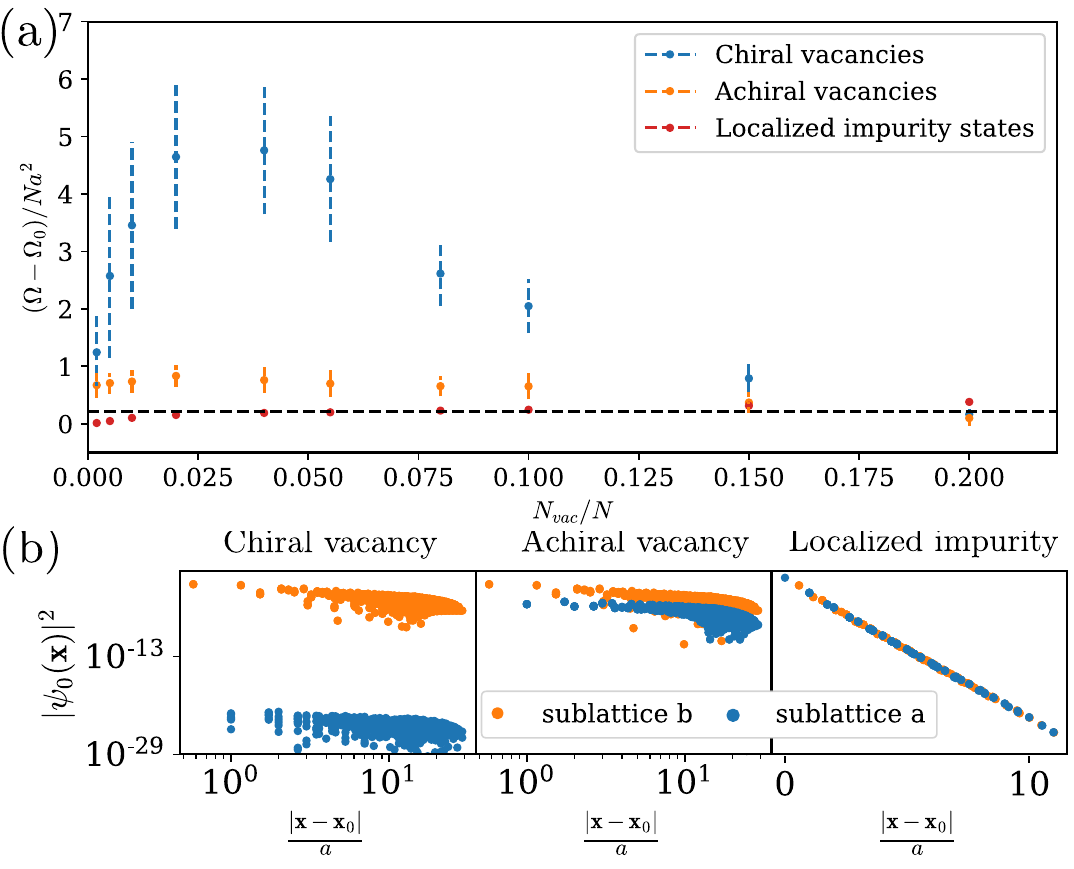}
    \caption{(a) Disorder induced quantum metric (per unit cell) as a function of defect concentration (beyond the pristine bulk value $\Omega_0$) for random distributions of vacancies (blue), achiral vacancies (orange), and localized defect states (red). Spread marks a standard deviation, extracted from averaging over 40 disorder configurations. Black dashed line shows $\Omega-\Omega_0 = S/2 = \sqrt{3}Na^2/8$, the expected enhancement for periodically spaced defects. (b) Squared amplitude of the electronic wave function as a function of the distance to the defect position for the three different types defects: the chiral vacancy as described earlier, the achiral vacancy and localized defect state obtained by adding an onsite potential $V = 4t$ to a single site. }
    \label{fig:density}
\end{figure}

In Fig.~\ref{fig:density}(a) we plot the enhancement of the quantum metric (beyond the pristine bulk value $\Omega_0$) at half-filling as a function of the vacancy concentration $n^* = N_{vac}/N$. The quantum metric (blue points) initially grows with defect concentration, but then saturates and eventually decreases.
To understand the small to moderate concentration regime, we imagine a situation where all vacancies are separated by the average distance $d^* = 1/\sqrt{n^*}$. 
An increase in concentration then results in more pairs contributing to the quantum metric, but with smaller contributions each, with the quantum metric enhancement given by $\Omega^* \propto \sum_{\mathrm{pairs\ i,j}} (\br_i-\br_j)^2 = \frac{Sn^*}{2}\times d^{*2} = \frac{S}{2}$, with $S$ the surface area of the sample times a constant.
We plot  $\Omega^*$ in Fig.~\ref{fig:density}(a) (dashed horizontal line) and note that the quantum metric is always substantially larger than this averaged value. Hence we conclude that ``rare" long-distance pairs must be responsible for the quantum metric increase with concentration. 
A similar phenomenology has been predicted in flat bands made of compact localized states~\cite{bouzerar2022}.
At a concentration $ \sim 5\%$, the effect of these ``rare" pairs saturates and they are instead superseded by closer neighboring pairs, eventually leading to neighboring vacancies, which explains the diminishing quantum metric at large concentrations.

\customSection{Other defects}\label{sec:other}
Finally, in order to further highlight the importance of chiral disorder and its related multifractal wave functions, we compare the quantum metric contributions from vacancies with related defects,
 obtained by instead adding an onsite potential $V$ to the defect site.
This creates two defect states with different characteristics.
One state has energy around $E = V$ and is exponentially localized to the defect site. We call this the localized defect state.  The other state remains close to zero energy, at $E\propto \frac{t}{1+V/t}$. 
It shares the same localization properties as the vacancy, except for the fact that it breaks chiral symmetry and then also has a finite weight on both sublattices. We therefore call this an achiral vacancy.
In the limit $V\rightarrow \infty$ the achiral vacancy becomes equivalent to the chiral vacancy state at $E=0$, as expected for hydrogenation, while the other, the localized defect state, is effectively removed from the spectrum.

In Fig.~\ref{fig:density}(b) we plot the localization profile as the square amplitude of the wave function for all three defect states: chiral vacancy, achiral vacancy, and localized defect state obtained for onsite defect potential $V = 4t$. 
We find a notable difference between the chiral and achiral vacancies when it comes to the sublattice occupation, directly tied to the chiral symmetry. Still, both are displaying the same long-range decay, tied to the multifractal nature of the chiral vacancy and clearly inherited by the achiral vacancy. This is in contrast to the localized defect which displays exponential localization. Finally we compare also the quantum metric for all three defects in Fig.~\ref{fig:density}(a). 
We find no enhancement of the quantum metric for localized defects. This is due to their exponentially decaying tail effectively introducing a cut-off distance, thus preventing hybridization over long distance and giving a vanishing quantum metric.
Achiral vacancies have a more interesting behavior. Despite their multifractal localization profile in Fig.~\ref{fig:density}(b), we find a notably smaller quantum metric enhancement than for chiral vacancies.
Thus, it is actually not only the multifractal wave nature that drives the enhanced quantum metric for vacancies, but preserving chiral symmetry is also crucial for an impurity band to generate a strongly enhanced quantum metric. 
This is consistent with the fact that hybridization does not occur when vacancies occupy the same sublattice, as established in Fig.~\ref{fig:dim}(d). 
The chiral symmetry also pins the system precisely at the critical point of the metal-insulator Anderson localization transition for all concentrations, with the quantum metric the key property to describe the special properties.
Breaking the chiral symmetry, the impurity band starts to localize and the system eventually becomes a trivial insulator with a quantum metric not influenced by the defects.

In summary, by introducing a space-resolved formulation of the quantum metric we show that random vacancies in a graphene monolayer generates strong long-range correlations at half filling, primarily 
from pairs of chiral symmetric vacancy states situated on different sublattices and with a small energy splitting, or equivalently, large spatial separation. Such ``rare" long-distance pairs provide an anomalous contribution, particularly at finite concentrations.
By utilizing similarities with the Anderson localization transition, we show that the multifractal vacancy wave functions is a key component of the quantum metric enhancement, but we also find that chiral symmetry is an additional important ingredient. A large quantum metric establishes the highly nontrivial nature of the flat vacancy band in graphene.
Possible extensions include studying the space-resolved quantum metric for other non-crystalline systems hosting critical states, including quasicrystals, fractals~\cite{iliasov2023}, or amorphous systems with vacancies. 
The spatial repartition of the quantum metric may also provide supplementary information about the electronic properties of the system. 
We also note that a recent work has discussed the relation between the quantum metric and the Berry curvature in the Haldane model in the presence of vacancies~\cite{romeral2024} and that some similar phenomenology was very recently also demonstrated in a Kitaev spin liquid~\cite{kao2023}.

\customSection{Acknowledgements}
We thank T. Löthman, R. Arouca, L. Baldo, and A. Bhattacharya for interesting and fruitful discussions related to this work. 
We acknowledge funding from the Knut and Alice Wallenberg Foundation through the Wallenberg Academy Fellows program and the European Research Council (ERC) under the European Union’s Horizon 2020 research and innovation programme (ERC-2022-CoG, Grant agreement No.~101087096).
Our calculations were performed using the Python package kwant~\cite{groth2014} and our plots using matplotlib~\cite{hunter2007}. The code used for the numerical calculations and the data shown in the manuscript is available at Ref.~\cite{marsal2023}.

\clearpage
\newpage


\section*{Supplementary material}

# Supplementary Material for "Enhanced quantum metric due to vacancies in graphene"

Quentin Marsal and Annica M. Black-Schaffer

*Department of physics and Astronomy, Uppsala University, Box 516, 751 20 Uppsala Sweden*

In this Supplementary Material (SM) we provide additional information supporting the derivations and results in the main text. First we provide additional information on the chiral symmetry of the system even with vacancies. Then we provide more details on some derivations of the quantum metric in the main text. Finally, we provide additional results showing that our conclusions are not dependent on system size.

## I. CHIRAL SYMMETRY

Since the Hamiltonian in Eq. (3) in the main text only contains nearest-neighbor hopping terms, it can be written in blocks as

$$\mathcal{H} = \begin{pmatrix} 0 & T \\ T & 0 \end{pmatrix}. \tag{1}$$

The presence of vacancies does not change this structure, but only the interior structure (terms) of $T$, since they only remove hopping terms. Thus, the Hamiltonian has sublattice, or chiral, symmetry: $\mathcal{S}^\dagger \mathcal{H} \mathcal{S} = -\mathcal{H}$ with $\mathcal{S} = \sigma_y \otimes I_{N/2}$, with $N$ being the number of sites in the system. We use a lattice built with armchair edges, to avoid spurious edge states at zero energy and thus it always contains an even number of sites. Due to the chiral symmetry, if $|\psi_\alpha\rangle$ is an eigenstate of the Hamiltonian with energy $E_\alpha$, then $|\psi_{\bar\alpha}\rangle = \mathcal{S}|\psi_\alpha\rangle$ is also an eigenstate with energy $E_{\bar\alpha} = -E_\alpha$. Figure 1(a) shows a histogram of the energy spectrum averaged over 20 lattice realizations. The accuracy of chiral symmetry can be estimated by plotting $E_{N/2+i} + E_{N/2-i}$ for $0 < i < N/2$, where the subscript labels the eigenvalues in increasing order. The result is shown in Fig. 1(b), which demonstrates that the only limit to chiral symmetry is the numerical precision of the diagonalization.

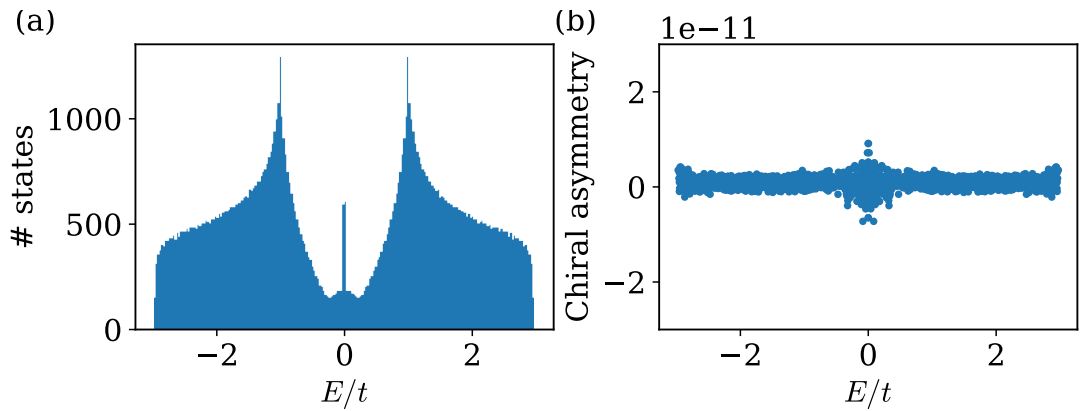

Figure 1. (a) Histogram of the energy spectrum for a system with 4% of vacancies, averaged for 20 lattice realizations. (b) Chiral asymmetry, computed as $(E_{N/2-i} + E_{N/2+i})/2t$, exemplified for one lattice realization. Note the multiplier of the $y$-axis.

## II. DERIVATION OF EXPRESSIONS FOR THE QUANTUM METRIC

The real space formulation of the quantum metric is introduced in Eq. (2) in the main text, here repeated for convenience:

$$\Omega_{\mu\nu} = \mathrm{Tr}\left[P \mathbf{r}_\mu (1-P) \mathbf{r}_\nu P\right], \tag{2}$$

In the case of a single pair of vacancies, if the bonding state $|\psi_+\rangle$ is occupied and the anti-bonding $|\psi_-\rangle$ is not, as expected at half-filling, then $P = |\psi_+\rangle\langle\psi_+|$ and $1 - P = |\psi_-\rangle\langle\psi_-|$. Hence, the quantum metric takes the following form:

$$\Omega_{pair} = \sum_\mu \text{Tr}[|\psi_+\rangle\langle\psi_+|\mathbf{r}_\mu|\psi_-\rangle\langle\psi_-|\mathbf{r}_\mu|\psi_+\rangle\langle\psi_+|] \tag{3}$$

$$= \sum_\mu |\langle\psi_+|\mathbf{r}_\mu|\psi_-\rangle|^2 \tag{4}$$

$$= \sum_\mu \left| \frac{\langle\psi_1| + \langle\psi_2|}{\sqrt{2}} \mathbf{r}_\mu \frac{|\psi_1\rangle - |\psi_2\rangle}{\sqrt{2}} \right|^2 \tag{5}$$

$$= \sum_\mu \frac{1}{4} |\langle\psi_1|\mathbf{r}_\mu|\psi_1\rangle - \langle\psi_1|\mathbf{r}_\mu|\psi_2\rangle + \langle\psi_2|\mathbf{r}_\mu|\psi_1\rangle - \langle\psi_2|\mathbf{r}_\mu|\psi_2\rangle|^2, \tag{6}$$

where $\psi_{1,2}(\mathbf{r}) = \psi_{vac}(\mathbf{r} - \mathbf{r}_{1,2})$ are single defect states, with most of their electronic weight around $\mathbf{r}_{1,2}$. For isotropic defect states with real wave functions, Eq. (6) reads

$$\Omega_{pair} = \sum_\mu \frac{1}{4} (\mathbf{r}_{1\mu} - \mathbf{r}_{2\mu})^2. \tag{7}$$

This expression is given as Eq. (4) in the main text. The single vacancy states of graphene are real but are not isotropic. Moreover, the bonding and anti-bonding states of the pair of vacancies are not an exact superposition of the single vacancy states, since each vacancy state scatters on the other vacancy. The exact value of the quantum metric therefore depends on the shape of the wave functions $|\psi_\pm\rangle$ and hence the results in Eq. (4) of the main text are an idealized result. Anisotropy steming both from the single vacancy states and from their pairing results in the spread of $\Omega_{pair}$ for each distance in Fig. 1(c) in the main text.

For a vacancy band, if we decompose $P$ and $1 - P$ into projectors $P_\alpha$ on individual eigenstates $|\psi_\alpha\rangle$, i.e.

$$P = \sum_{\alpha \text{ filled}} |\psi_\alpha\rangle\langle\psi_\alpha| = \sum_{\alpha \text{ filled}} P_\alpha, \tag{8}$$

and

$$1 - P = \sum_{\beta \text{ empty}} |\psi_\beta\rangle\langle\psi_\beta| = \sum_{\beta \text{ empty}} P_\beta, \tag{9}$$

then the total quantum metric reads

$$\Omega = \sum_\mu \sum_{\alpha \text{ filled}} \sum_{\beta \text{ empty}} \text{Tr}[P_\alpha \mathbf{r}_\mu P_\beta \mathbf{r}_\mu P_\alpha]. \tag{10}$$

This expression is given as Eq. (5) in the main text. We note that each term is here necessarily positive, since

$$\text{Tr}[P_\alpha \mathbf{r}_\mu P_\beta \mathbf{r}_\mu P_\alpha] = \text{Tr}[P_\alpha \mathbf{r}_\mu P_\beta P_\beta \mathbf{r}_\mu P_\alpha] \tag{11}$$

$$= \text{Tr}[P_\alpha \mathbf{r}_\mu P_\beta (P_\alpha \mathbf{r}_\mu P_\beta)^\dagger]. \tag{12}$$

### III. SCALING WITH SYSTEM SIZE

All numerical results in the main text are computed on finite sized lattices. To illustrate the effect of having a finite lattice on the quantum metric, we show on Figure 2 the same data as Fig. 1(c) in the main text but for different system sizes. Note that the quantum metric is here for simplicity plotted as a function of the squared distance between vacancies. The figure shows how the finite size of the system affects the pristine value of the quantum metric (i.e. the contribution from bulk states) and the numerical prefactor, but that it, notably, has no influence on the quadratic dependence in the distance between vacancies. The system size also has no effect the spreading of the quantum metric for a given distance, but that is instead created by the anisotropy of the lattice and thus the vacancy states, as discussed in the main text. We also note that for large vacancy pair distances the results converge for all system sizes. These results shows that system size is not influencing our results in Fig. 1. We have checked for similar finite size effects for all our results.

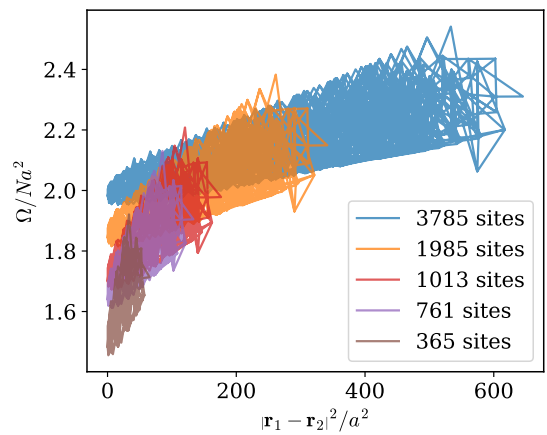

Figure 2. Quantum metric generated by a single pair of vacancies laying on different sublattices, as a function of the squared distance between the vacancies.


%
%
\end{document}